# Plasmon-Mediated Hybridization of Wannier-Mott and Frenkel Excitons in a Monolayer $WS_2$ - J-Aggregate Hybrid System

Nicolas Zorn Morales, Daniel Steffen Rühl, Sergey Sadofev, Emil List-Kratochvil, and Sylke Blumstengel*

Sylke Blumstengel
Department of Physics, Department of Chemistry & Center for the Science of Materials Berlin, Humboldt-Universität zu Berlin, Zum Großen Windkanal 2, 12489 Berlin, Germany; orcid.org/0000-0002-1478-9757; Email: sylke.blumstengel@physik.hu-berlin.de

Nicolas Zorn Morales
Department of Physics, Department of Chemistry & Center for the Science of Materials Berlin, Humboldt-Universität zu Berlin, Zum Großen Windkanal 2, 12489 Berlin, Germany; orcid.org/0000-0002-5910-8484

Daniel Steffen Rühl
Department of Physics, Department of Chemistry & Center for the Science of Materials Berlin, Humboldt-Universität zu Berlin, Zum Großen Windkanal 2, 12489 Berlin, Germany; orcid.org/0000-0001-8373-0102

Sergey Sadofev
Leibniz Institute of Crystal Growth, Max-Born-Straße 2, 12489 Berlin, Germany

Emil J.W. List-Kratochvil
Department of Physics, Department of Chemistry & Center for the Science of Materials Berlin, Humboldt-Universität zu Berlin, Zum Großen Windkanal 2, 12489 Berlin, Germany and Helmholtz-Zentrum Berlin für Materialien und Energie GmbH, Hahn-Meitner-Platz 1, 14109 Berlin, Germany; orcid.org/0000-0001-9206-800X

We present a tunable plasmonic platform that allows room temperature hybridization of dissimilar excitons, namely of Wannier-Mott excitons in monolayer (1L) $WS_2$ and Frenkel excitons in molecular J-aggregates via simultaneous strong coupling to surface plasmon polaritons. It is based on a simple layered design consisting of a thin planar silver film and a dielectric spacer on which monolayer and the aggregates are assembled. Strong coupling is revealed by angle-dependent spectroscopic ellipsometry measurements in total internal reflection geometry by the observation of double Rabi splitting at the two excitonic resonances. We analyze the exciton-exciton-plasmon system with the coupled oscillator model and demonstrate modulation of polariton character and dynamics by the number of molecules participating in the coupling. Furthermore, we propose a route to remotely control the mode splitting at the Frenkel excitonic resonance via electrostatic gating of the 1L-$WS_2$ and to switch the molecule-plasmon interaction between the weak and strong coupling regime.

## 1. Introduction

Hybrid light-matter excited states - also called polaritons - have attracted intense interest over the past decades due their intriguing physics arising from the blending of properties of light and matter and the prospect of applications in various fields of physics and chemistry. While research interest was initially triggered by applications ranging from low-threshold lasing in semiconductors to photon quantum information,[1] in more recent years, the possibility to manipulate matter properties, like charge and energy transport in organic materials, and chemical reactivity was discovered and is now intensively studied.[2] Ingredients for creation of hybrid light-matter states are intense narrow-band electronic or vibrational transitions in matter and a reduction of the effective mode volume for photons to achieve sufficiently large light-matter coupling strengths, a prerequisite to reach the strong coupling (SC) regime. Surface plasmon polaritons (SPP), i.e. electromagnetic excitations propagating at the interface between dielectric and metal, allow for strong electromagnetic field confinement and local field enhancement.[3] On the matter side, 2D transition metal dichalcogenides (TMDC)[4, 5] and molecular J-aggregates are particularly suited since they both feature narrow excitonic transitions with huge oscillator strengths and large exciton binding energies so that excitons persist up to room temperature. Consequently, SC between excitons in TMDCs[6, 7] or J-aggregates[8] with SPPs supported by plasmonic nanoparticles, plasmonic nanocavities or thin metal films has been widely studied. Besides room-temperature operation, the feasibility of active control of the light-matter coupling strength via electrostatic gating[9, 10], temperature[11, 12], fs optical pumping[13] in plasmonic structures with TMDCs and via the number of oscillators present in molecule-based plasmonic systems[14] is particularly appealing. It is straightforward to extend these studies and to attempt to construct a plasmonic platform where the SPP not just couples with one exciton species but mixes excitons of different materials, namely Wannier-Mott (WM) excitons in the TMDC and Frenkel (F) excitons of the J-aggregate to yield a hybrid light-matter state containing excitons of very different nature.

Multimode coupling involving either different exciton species or plasmonic modes[15] introduces novel aspects. In such case, the plasmonic or photonic nanocavity can sustain multiple coherent states and has, as such, more energy dissipation channels and greater parameter space for modulation. Early works, attempting to achieve exciton-exciton-photon hybrid states by integration of epitaxial quantum dots in photonic crystals,[16] were trigger by the prospect to achieve simultaneous strong coupling with a nanocavity at the single photon level which is very favorable for quantum entanglement generation.[17] However, multimode coupling was only observed at low temperature (<10K) since excitons in the traditional III-V semiconductors are not stable at room temperature. With the new materials at hand, room temperature operation comes into reach.[18] For example, SPP coupling with different exciton species in a single material, namely $WS_2$, has been observed involving either excitons and trions (at low temperature)[19] or A and B excitons.[20] On the other hand, SPP-mediated coupling of excitons residing in different materials is appealing also with respect to energy transport. The hybridization is predicted to yield enhanced energy transfer between both materials, and thanks to the intrinsic delocalized character of the plasmonic modes, exciton transport at macroscopic length scales can be achieved. [21, 22] With a similar motivation, also photonic cavities have been studied where the cavity photons mediate the coupling between different exciton species. [23]

In this contribution we propose a very simple planar plasmonic platform for achieving hybrid light-matter excited states involving Wannier-Mott excitons (WMX) in 1L-$WS_2$ and Frenkel excitons (FX) of dye aggregates of 5,6-dichloro-2-[[5,6-dichloro-1-ethyl-3-(4-sulfobutyl)-benzimidazol-2-ylidene]-propenyl]-1-ethyl-3-(4-sulfobutyl)-benzimidazolium hydroxide, inner salt, sodium salt (TDBC). TDBC is a thoroughly studied cyanine dye which shows J-aggregation both in solution and in solid state.[24] We employ total internal reflection ellipsometry (TIRE) which combines spectroscopic ellipsometry with the Kretschmann-Raether-type surface plasmon resonance (SPR) geometry.[7, 25] Via angle-resolved measurements we reveal the formation of SPP - FX - WMX hybrid states. Applying the coupled oscillator model allows us to derive the coupling strengths and Hopfield coefficients. We furthermore explore degrees of freedom offered to control polariton dispersion, polariton composition and dynamics.

## 2. Results

The plasmonic platform is depicted in **Figure 1a**. It consists of a 2D hybrid stack comprised of a silver film, a thin $Al_2O_3$ spacer, 1L-$WS_2$ and a thin layer of TDBC J-aggregates in a polyvinyl alcohol matrix on a BK7 glass slide. Large area 1L-$WS_2$ (1 cm$^2$) was obtained by pulsed thermal deposition (PTD)[26] and subsequent polymer-based wet-transfer. The $Al_2O_3$ interlayer was introduced to protect the Ag film during the wet-transfer process and to avoid direct contact of 1L-$WS_2$ with the metal film which is known to deteriorate its optical properties. The chosen Ag film thickness assures strong and narrow plasmon resonances at around 2 eV which is essential for achieving the SC regime.[14] We investigate two hybrid stacks termed A and B which differ in TDBC concentration and film thickness.

Before presenting the results, we recollect the TIRE principle (see Figure 1b).[25] An ellipsometry measurement yields the ellipsometric angles Δ and Ψ which are related to the ratio of the complex Fresnel reflection coefficients $r_p$ and $r_s$ for *p*- and *s*-polarized light, respectively, via the equation $\rho = \frac{r_p}{r_s} = \tan(\Psi) \cdot e^{i\Delta}$. $\tan(\Psi) = |r_p/r_s|$ is the amplitude of $\rho$ while $\Delta = \varphi_p - \varphi_s$ is the phase difference between the reflection coefficients of *p*- and *s*-polarized light. The rather unintuitive ellipsometric angles can be better understood remembering that SPPs at a metal-dielectric interface can only be excited with *p*-polarized light [27]. Ellipsometry yields thus the amplitude and phase spectra of the polarization experiencing the resonance relative to the orthogonal polarization which experiences no resonance. The SPP resonance manifests itself as a minimum in the amplitude $|r_p|$ as well as a rapid change in phase $\varphi_p$. Energy and momentum conservation determine thereby the resonance energy of the SPP for a given angle of incidence $\theta$. Since *s*-polarized light does not excite SPPs, for the reflection coefficient holds $|r_s| \approx 1$ in the considered spectral range. Consequently, $\tan(\Psi) \approx |r_p|$ simply represents the reflection coefficient spectrum in *p*-polarization. The minima in the Ψ spectra correspond to the energies of the three coupled SPP-exciton-exciton resonances. Angle-dependent measurements of Ψ spectra allow thus the construction of the polariton dispersion. The plot $1 - \tan^2(\Psi)$ corresponds to the absorption spectrum.

Figure 1c introduces the three individual components. Presented are the absorption coefficient spectra of TDBC:PVA and 1L-WS$_2$ in the hybrid stack. The spectra are derived from ellipsometry measurements (supporting information, section S1). The TDBC spectrum with a maximum at 2.095 eV shows the typical signature of J-aggregation. While the TDBC monomer absorption lies at 2.36 eV, J aggregation causes a red shift of the absorption maximum.[24] The TDBC:PVA film in hybrid stack A has a $\kappa_{max} = 0.25$ and hybrid stack B $\kappa_{max} = 0.4$. The 1L-WS$_2$ absorption with a maximum at 2.02 eV is due to the A excitonic transition.[5] The absorption coefficient $\kappa$ is smaller than that typically reported in the literature.[28] This might be related to the wet transfer process which deteriorates somewhat the quality of our PTD-grown monolayer. Here, one should bear in mind that the beam spot diameter of the ellipsometer is rather large (ca. 1 mm) and therefore a large area of the 1L-WS$_2$ is interrogated. The detuning between the FX and WMX resonances is 75 meV, which is larger than the individual linewidths, which are $\gamma_{WMX} = 54$ meV for 1L-WS$_2$ and $\gamma_{FX} = 42 \pm 2$ meV for TDBC:PVA. Also depicted in Figure 1c is the spectrum of the third ingredient, namely the SPP of a pristine Ag film tuned to the center position between the FX and WMX resonances. It is obtained from a TIRE measurement on a sample spot not covered by 1L-WS$_2$ and prior deposition of TDBC:PVA. Due to the strong damping in the metal, SPP line width $\gamma_{SPP} = 144$ meV is considerably larger than that of the excitons. Note that the resonance energy of the SPP for a given angle of incidence $\theta$ or a given in-plane wavevector $k_\parallel = \frac{2\pi \cdot n_{BK7}}{\lambda} \sin\theta$ with $n_{BK7}$ being the refractive index of the prism and $\lambda$ the wavelength of the incoming photons, is determined by energy and momentum conservation. Via variation of $\theta$ the SPR can thus be tuned across the WMX and FX resonances and this is employed to study the coupling of the SPP-WMX-FX system.

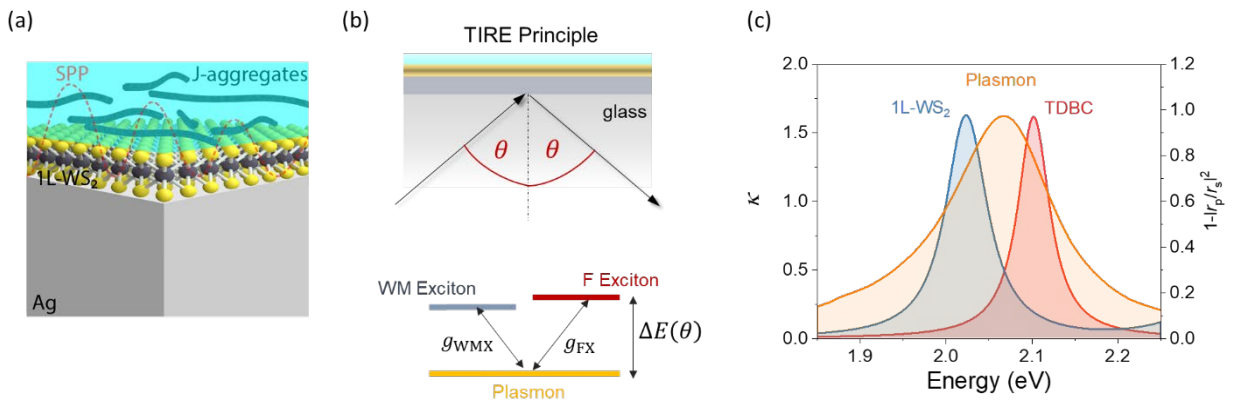

**Figure 1.** a) Sample layout to study the coupling between SPPs of an Ag film (grey) and WM excitons of 1L-WS$_2$ and F excitons of TDBC J-aggregates in a PVA matrix (turquoise). The investigated hybrid stacks consist of Ag (51 nm)/Al$_2$O$_3$ (2 nm)/1L-WS$_2$ /TDBC:PVA on BK7

glass. The TDBC:PVA film thicknesses are 12 nm in hybrid stack A and 17.5 nm in hybrid stack B. The Al$_2$O$_3$ layer is omitted in the figure. b) TIRE geometry to tune the SPR and thus the energy detuning $\Delta E$ via the angle $\theta$ of the incident light beam. c) Absorption coefficient $\kappa$ spectra of 1L-WS$_2$ (blue) and TDBC (red). The latter is normalized to $\kappa_{\max}$ of the TMDC. Also depicted is an absorption spectrum of the SPP of a bare Ag/ Al$_2$O$_3$ stack (orange). More spectra are provided in the supporting information, section S2.

**Figure 2a** shows $\Psi$ spectra obtained from angle-dependent TIRE of the hybrid stack A. The three minima visible are due to photoexcitation of the three coupled SPP-WMX-FX resonances. The plasmon resonance, clearly discernable at small and large $\theta$, blueshifts with increasing angle thereby experiencing anticrossing when approaching the WMX as well as the FX resonance, which is a strong indication for SC. This is better seen in Figures 2b and c displaying the corresponding density plots of $|r_p/r_s|$ as a function of photon energy and $\theta$. The plots are constructed from respective TIRE measurements of hybrid stack A and B (supporting information, Figure S3). Now clearly visible are three polariton branches with an avoided crossing and Rabi splitting evident at both, the 1L-WS$_2$ WMX and the TDBC FX energy, which we term in the following lower, middle and upper polariton branch (LPB, MPB, UPB). The difference in the Rabi splittings $\Omega$ at the FX resonance between hybrid stack A and B reflects the difference in the number $N_m$ of TDBC molecules participating in the coupling, as detailed below. To gain more inside into the coupling and the nature of the three hybrid polariton states we describe our system with a coupled oscillator model where the SPP mode simultaneously couples to the two excitonic transitions with coupling constants $g_k$ (k = WMX, FX). Direct coupling of WMX and FX is described by the coupling constant $g_X$. In the basis of the uncoupled $|SPP\rangle$, $|WMX\rangle$ and $|FX\rangle$ modes, the Hamiltonian reads:

$$\begin{pmatrix} E_{SPP} + i\gamma_{SPP} & g_{WMX} & g_{FX} \\ g_{WMX} & E_{WMX} + i\gamma_{WMX} & g_X \\ g_{FX} & g_X & E_{FX} + i\gamma_{FX} \end{pmatrix}. \qquad (1)$$

Here, $E_j$ (j = SPP, WMX, FX) are the energies of the three uncoupled modes. In our hybrid stacks, the TDBC aggregates are highly diluted in a PVA matrix with a TDBC:PVA monomer ratio of $(2\text{-}3)\cdot 10^{-3}$. The film thickness is 12 nm and 17.5 nm, respectively. Since in average the molecules are not in close proximity to the 1L-WS$_2$ surface, we neglect for the moment direct excitonic coupling (i.e. $g_X = 0$) and explore its effect on the polariton dispersion later.

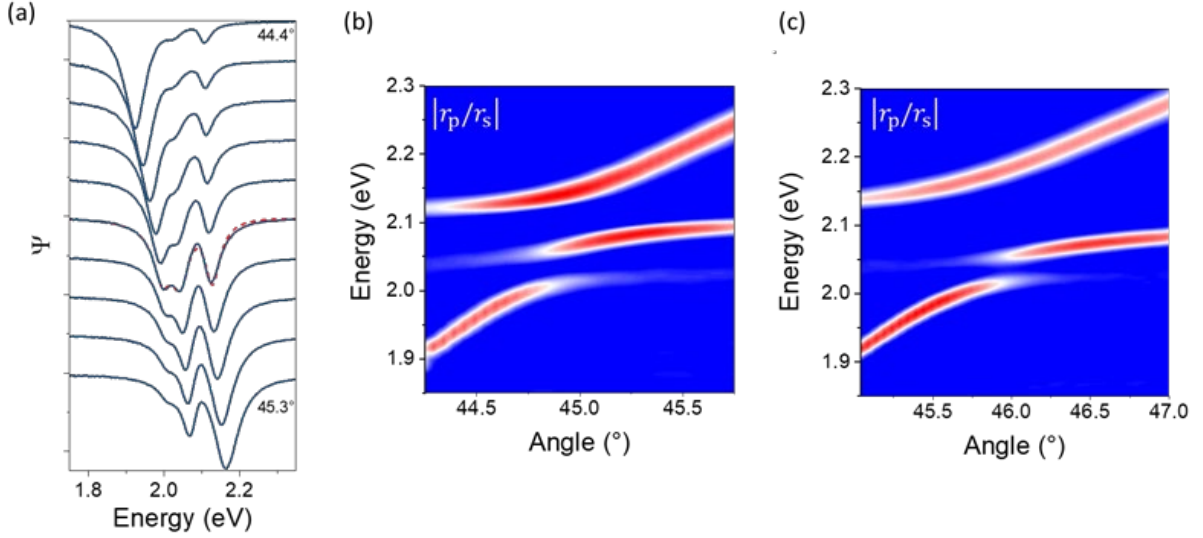

**Figure 2.** a) Angle-dependent Ψ spectra of hybrid stack A. The red dashed line shows a fit of the spectrum at 44.9° to derive the absorption coefficient spectra depicted in Fig. 1c. For details see the supporting information. Density plots of the $|r_p/r_s|$ spectra vs. photon energy and incident angle constructed from TIRE data of hybrid stacks A (b) and B (c). For better contrast, the second derivative $\frac{\partial^2}{\partial E^2}|r_p/r_s|$ is plotted.

Fitting of the dispersion curves plotted versus the wavevector $k_\parallel$ (**Figures 3a and b**) with the known values for energies and broadening parameters and setting $g_X = 0$ yields Rabi splittings $\Omega_k = 2g_k$ of $\Omega_{WMX} = 51$ meV and $\Omega_{FX} = 60$ meV for hybrid stack A and of $\Omega_{WMX} = 51$ meV and $\Omega_{FX} = 76$ meV for hybrid stack B. The obtained $\Omega_{WMX}$ agree well to previous findings[7] supporting the assumption of negligible direct WMX-FX coupling. A more thorough justification of this approximation is provided in the supporting information, section S4. In the present configuration, the coupling strength $g_k$ is proportional to $\sqrt{E_k \gamma_k \kappa_k d_k / L_z}$, with $d_k$ being the layer thicknesses of 1L-WS$_2$ and TDBC:PVA, respectively, and $\sqrt{\gamma_{FX} \kappa_{FX} d_{FX}} \sim \sqrt{N_m}$.[7] $L_z$ is the effective mode length of the SPP which is a measure of the extension of the electric field $E$ of the SPP into the dielectric which falls off exponentially perpendicular to the interface. A common definition of $L_z$ is $L_z w_{E,max} = \int_{-\infty}^{\infty} w_E dz$ with $w_E$ being the electric field energy density.[27] For a pristine Ag surface in air $L_z \approx 225$ nm at 2 eV is obtained. In the present planar configuration $d_k \ll L_z$ holds explaining the generally small Rabi splittings. It should be noted, that we have intentionally prepared very thin and highly diluted TDBC:PVA films to achieve splittings of the same order of magnitude as for the atomically thin 1L-WS$_2$. As pointed out above, the quality of our 1L-WS$_2$ layer over the large area interrogated is not perfect which manifests itself in a $\kappa_{\text{WMX}}$ value somewhat lower than typically observed. Using literature data[28] for 1L-WS$_2$ it should be possible to achieve a coupling strength about 1.4 times larger. Nevertheless, for both types of excitons, the derived values fulfill the SC criterium $\hbar\Omega_k > |\gamma_k - \gamma_{\text{SPP}}|/2$ which is equivalent that at resonance, i.e. $E_k = E_{\text{SPP}}$, the complex eigenenergies exhibit a splitting in their real parts [Re($E$)] while the imaginary parts [Im($E$)] remain unsplit.[29] The latter means that at resonance, the polariton branches decay with a common rate. This is indeed the case as seen in Figures 3c and d which show crossings of the dissipation rates of the three polariton branches when plotted as a function of the wavevector. Furthermore, the Rabi splitting is larger than the half width at half maximum (HWHM) of the coupled resonances, i.e. $\frac{\gamma_k + \gamma_{\text{SPP}}}{4}$, which might be considered a more intuitive criterium for SC. The eigenfunctions pertaining to the three polariton branches are quantum superpositions of the uncoupled SPP, WMX and FX. Their respective contributions are proportional to the square modulus of the respective eigenvector coefficients $|\alpha_j|^2$ (Hopfield coefficients). In Figures 3e and f we plot the $|\alpha_j|^2$ for the most interesting MPB. As apparent, it changes from WM excitonic to F excitonic nature for increasing wavevector. At intermediate wavevectors, the MPB contains contributions from both the WM and F excitons indicating that the two types of excitons are hybridized and that the hybridization is mediated by the SPP. While in hybrid stack A the plasmonic part dominates, in hybrid stack B the fraction of excitons is larger. The weight of the total excitonic contributions can consequently be controlled by the number of molecules $N_m$ participating in the coupling. The transition from more plasmon-like to a more exciton-like state results in an alteration of the dissipation rate of the MPB (Figures 3c and d) and thus of the emission dynamics. Furthermore, the character of the excitonic part can be tailored by the number of molecules $N_m$ in particular in the LPB. At intermediate wavevectors, it can be tuned from purely WMX-like to dominantly FX-like just be changing $N_m$ (supporting information, Figure S5).

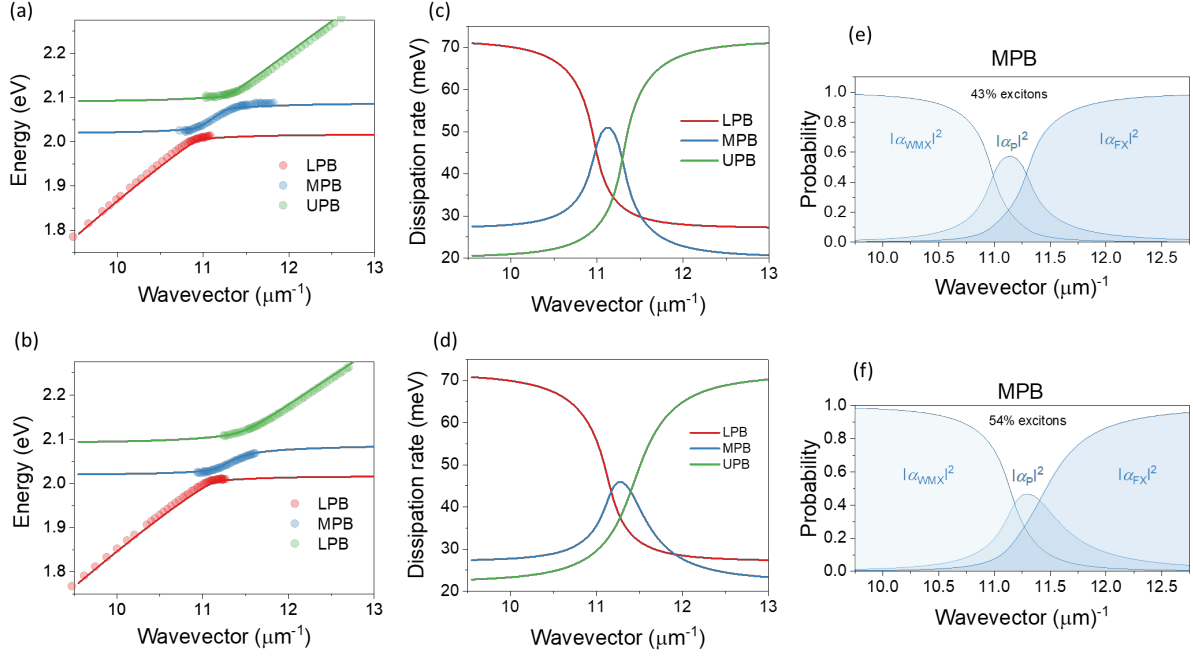

**Figure 3**. Dispersion relation $E$ vs. $k_\parallel$ of the coupled WMX-FX-SPP resonances of hybrid stack A (a) and B (b). The symbols represent the energetic positions of the minima in the $|r_\mathrm{p}/r_\mathrm{s}|$ spectra. The solid line is a fit to the data using the coupled oscillator model (Equation 1) as described in the main text. The obtained coupling strengths are $g_\mathrm{WMX} = 25.5$ meV and $g_\mathrm{FX} = 30$ meV for hybrid stack A and $g_\mathrm{WMX} = 25.5$ meV and $g_\mathrm{FX} = 38$ meV for hybrid stack B. Derived dissipation rates [Im($E$)] of the three polariton branches for hybrid stack A (c) and B (d). Hopfield coefficients of the MPB for hybrid stack A (e) and B (f).

In our above discussion, we neglected direct coupling between F and WM excitons which is not always justified. In a previous work it was shown that energy transfer from a molecule (PTCDA) to 1L-MoS$_2$ takes place on a sub-ps time scale when donor and acceptor are in close proximity[30] pointing at a substantial transition matrix element $M_\mathrm{if}$. A rough estimate of $M_\mathrm{if}$ applying Fermi's golden rule $k_\mathrm{ET} = \frac{1}{h}|M_\mathrm{if}|^2 \rho_f$ yields ~10 meV with $k_\mathrm{ET} \approx 1$ ps$^{-1}$ and assuming a density of final states $\rho_f \approx 2/\pi\gamma_\mathrm{WMX}$. For a Förster-type energy transfer and at small distances, the rate shows a combination of exponential and power law decay as a function of the distance between molecules and TMDC.[31] Therefore, Förster-transfer is efficient over very small distances (a few nm) only. Since it is feasible to either directly grow J-aggregates on TMDC surfaces or contrarily to transfer TMDC monolayers on J-aggregates we explore in the following the effect of an additional WMX-FX coupling. A question of particular interest in terms of practical applications is if via participation of a second excitonic species the exciton-SPP coupling of the other species can be influenced. To this end, we consider a scenario where the TDBC aggregate concentration is further reduced so that the coupling with the SPP becomes weak.

**Figures 4a and b** show the dispersion of the real and imaginary parts of the complex Eigenenergies $E_\pm$ of the coupled SPP-FX system which is obtained by setting $g_{WMX} = 0$. Here, the $g_{FX}$ is chosen such that the condition for weak coupling, namely $\hbar\Omega_R < |\gamma_{FX} - \gamma_{SPP}|/2$ holds. Consequently, the real parts of $E_\pm$ coalesce at resonance (Figure 4a) while simultaneously a splitting in the imaginary parts arises which implies a modification of the decay rate of the FX (Figure 4b). Now we add 1L-WS$_2$ and introduce direct WMX-FX coupling. Note that the WMX-FX coupling constant (12 meV) is smaller than the detuning (75 meV) between the resonances. The respective dispersion of the real parts of the Eigenenergies is depicted in Figure 4c. Now, an avoided crossing at both excitonic resonances is clearly observed. This means, that 1L-WS$_2$ WMX coupled to the SPP, by providing oscillator strength to the molecules, augments the coupling of the J-aggregates with the SPP and drives it even into the SC regime. This is also seen by plotting the Hopfield coefficients for the three polariton branches depicted in Figure 4d. The MPB changes again from WM excitonic to F excitonic character with increasing wavevector, while at intermediate wavevectors, superposition of all three constituents is observed with a large exciton contribution (47%) indicating that the two types of excitons are now strongly hybridized. Also the UPB still contains a significant contribution of the WMX (11%) at intermediate wavevector due to the additional WMX-FX coupling.

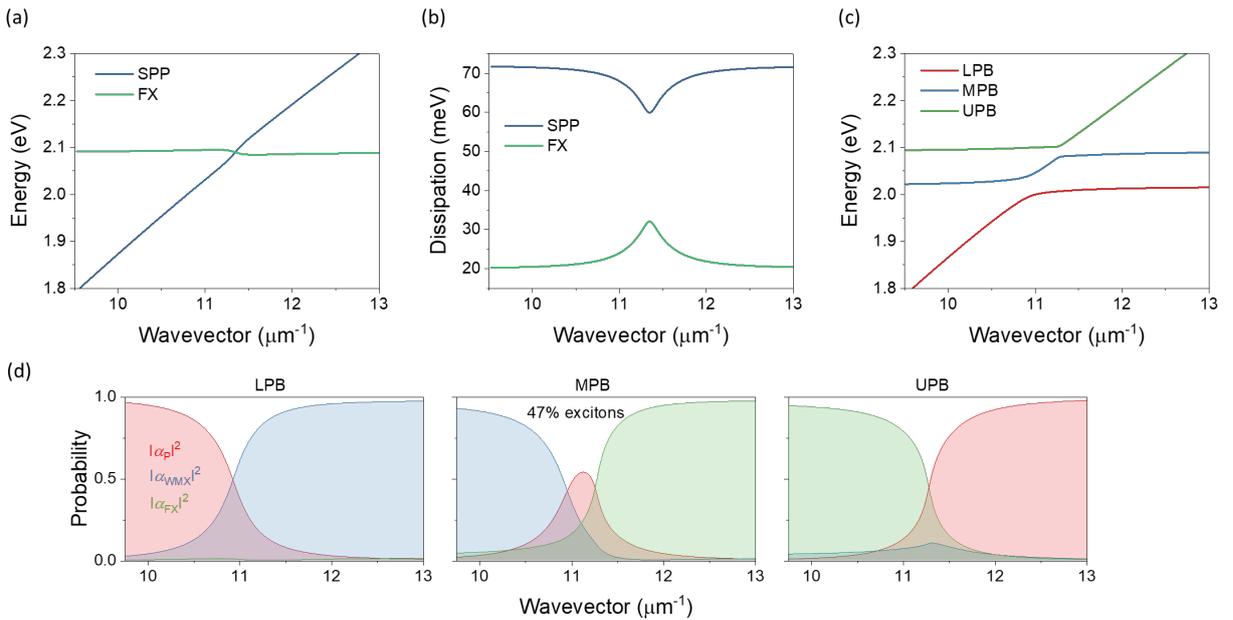

**Figure 4**. Energy dispersion (a) and dissipation rate (b) for a hybrid stack without 1L-WS$_2$. The FX-SPP coupling strength is reduced to $g_{FX} = 22$ meV, so that the system is in the weak coupling regime. All other parameters are left unchanged. c) Energy dispersion of a full hybrid stack with 1L-WS$_2$. Now, direct coupling between FX and WMX is introduced in the coupled oscillator model with a coupling strength of $g_X = 12$ meV. $g_{FX} = 22$ meV and $g_{WMX} = 35$ meV. The latter is reasonable with reported optical constants of high quality 1L-WS$_2$.[28] All other parameters are left unchanged. d) Hopfield coefficients for the three polariton branches for the dispersion relation plotted in (c).

The finding provides a simple means for external control the exciton-plasmon coupling of molecules. Since the absorption coefficient of 1L-WS$_2$ and thus $g_{WMX}$ can be strongly modified by external means, for example, by electrostatic gating, the participation of the TMDC in the coupling can be switched off and on and, by doing so, it is possible to control the mode splitting at the FX resonance and to drive the molecules even between the weak and strong coupling regimes. This facilitates, for example, the switching of long-distance plasmon-mediated energy transport in a donor-acceptor system separated laterally on μm length scale.[22] The Ag/Al$_2$O$_3$/1L-WS$_2$ hybrid stack could provide a suitable platform to assemble such donor-acceptor system atop whereby the metal/insulator/semiconductor configuration simultaneously allows for electrostatic gating. Furthermore, the prospect of modification of chemical reactivity under SC has recently been recognized.[2] Assembly of molecules on the Ag/Al$_2$O$_3$/1L-WS$_2$ hybrid stack could allow study of surface chemical and molecular processes under the conditions of weak and strong coupling with confined electromagnetic fields. On the other hand, due to the atomic thinness of 1L-TMDCs, the coupling constant $g_{WMX}$ is notoriously small. Switching roles, off-resonant molecules could increase the level splitting at the WMX resonance and provide an alternative route to render SC with 1L-TMDCs more robust which is very much desired for the construction of plasmonic modulators, switches as well as for applications based on the creation of polariton condensates that require large splitting-to-linewidth ratios up to room temperature.[9, 11]

## 3. Conclusion

In conclusion, we have demonstrated a simple plasmonic platform that allows hybridization of excitons of very dissimilar materials, namely of those of a molecular J-aggregate and those of 1L-WS$_2$. The participation of two exciton species provides more control over energy dissipation channels and large parameter space for modulation. Via changing the number of molecules participating in the coupling, the character of the polariton mode can be changed from more photon- to more exciton-like and consequently the dissipation rate controlled. The design of the platform provides furthermore remote control of the coupling strength of the 1L-TMDC WMX via electrostatic gating. This becomes particularly appealing when direct coupling between the FX- and WMX-resonances is present. In this case, the observed mode splitting at the aggregates FX resonance can be remotely controlled via gating of the TMDC and the interaction switched between the weak and strong coupling regimes.

## 4. Experimental Section/Methods

*Sample preparation*: 1L-WS$_2$ was grown by pulsed thermal deposition on sodalime glass.[26] The large area (1 cm$^2$) 1L-WS$_2$ was placed by a polymer-based wet-transfer process on a ~50 nm thick Ag film thermally evaporated on BK7 glass and covered by a 2 nm thick Al$_2$O$_3$ spacer layer fabricated by atomic layer deposition. TDBC (FEW Chemicals) thin films in a polyvinyl alcohol matrix were prepared by spin coating. A stock solution of 1.2mg/ml TDBC in de-ionized water was prepared and left for at least 24h for J-aggregate formation. To adjust the TDBC concentration, the stock solution was diluted just immediately prior film preparation which involved mixing with a PVA solution in deionized water and spin coating. This procedure prevents dissolution of J-aggregates into monomers.

*Ellipsometry*: The glass slide was fixed on a cylinder prism by an index matching oil to realize excitation of SPPs in the Ag film in the Kretschmann-Raether configuration. The angle of incidence $\theta$ is set with an accuracy of $\pm 0.5°$ in the experiments. The beam spot diameter of ellipsometer (SENresearch 4.0, Sentech) is ca. 1 mm.

**Acknowledgements**

NZM, DSR, ELK and SB gratefully acknowledge financial support by the Deutsche Forschungsgemeinschaft through CRC 951 (Project number 182087777).

# Supporting Information

**Plasmon-Mediated Hybridization of Wannier-Mott and Frenkel Excitons in a Monolayer WS₂ - J-Aggregate Hybrid System**

Nicolas Zorn Morales, Daniel Steffen Rühl, Sergey Sadofev, Emil List-Kratochvil, and Sylke Blumstengel*

**S1. Spectroscopic ellipsometry of Ag/Al₂O₃/1L-WS₂/TDBC:PVA stacks to determine the dielectric functions of 1L-WS₂ and TDBC:PVA**

The dielectric functions and film thicknesses of 1L-WS₂ and TDBC:PVA in the hybrid stacks A and B are obtained by fitting the ellipsometric spectra Ψ and Δ at the given angle of incidence (**Figure S1**).

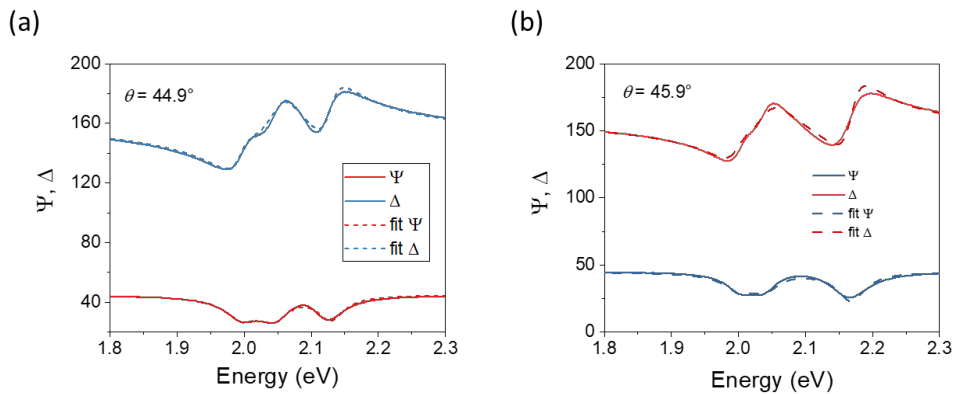

**Figure S1.** Ψ (red) and and Δ (blue) spectra at the given angle of incidence $\theta$ of hybrid stack A (a) and B (b) attached to a BK7 glass prism.

Hereby, we use literature data for the dielectric functions of BK7 glass and Al₂O₃.[1] The dielectric function and the thickness of the Ag film were determined by an ellipsometric measurement before deposition of 1L-WS₂ and TDBC:PVA. The A, B, C and D excitonic transitions in 1L-WS₂ are modelled by Tauc-Lorentz oscillators. The imaginary part of the susceptibility $\chi_i$ of a single oscillator is given by:

$$\mathrm{Im}(\chi_i) = \begin{cases} \dfrac{1}{E} \cdot \dfrac{A_i E_{0,i} C_i (E - E_{g,i})^2}{(E^2 - E_{0,i}^2)^2 + C_i^2 E^2} & \forall\ E > E_{g,i} \\ 0 & \forall\ E < E_{g,i} \end{cases},$$

The real part of $\chi$ is obtained by Kramers-Kronig transformation. The dielectric function of 1L-$WS_2$ is given by $\varepsilon_{WS_2} = \varepsilon_\infty + \sum_i \chi_i$. As starting values for the parameters of the Tauc-Lorentz oscillators served literature data.[2] Since at the given angle of incidence the spectral response of the ellipsometric angles Δ and Ψ is flat in the spectral range of the B, C, and D excitonic transitions, the fit is only sensitive to the parameters of the A exciton and the layer thickness. The extracted parameters are given in Table S1.

Table S1: Tauc-Lorentz parameters for ML-$WS_2$ ($\varepsilon_\infty$ =4.56)

|  | $A$ (eV) | $E_0$ (eV) | $C$ (meV) | $E_g$ (eV) |
|---|---|---|---|---|
| A-exciton | 1.873 | 2.018 | 54.5 | 0.77 |
| B-exciton | 109 | 2.383 | 211 | 2.11 |
| C-exciton | 38 | 2.837 | 213 | 2.28 |
| D-exciton | 237 | 2.901 | 889 | 2.28 |

The dielectric function of TDBC:PVA is modelled with a single Lorentz oscillator. The background dielectric function is fixed to the value of PVA. The extracted absorption coefficient spectra of 1L-$WS_2$ and TDBC:PVA as well as the parameters of the Lorentz oscillator are given in the main manuscript.

**S2. Ψ spectra of Ag/$Al_2O_3$ at selected angles of incidence**

Figure S2 shows ellipsometric spectra Ψ of an Ag/$Al_2O_3$ stack attached to a BK7 glass prism recorded at three different angles of incidence.

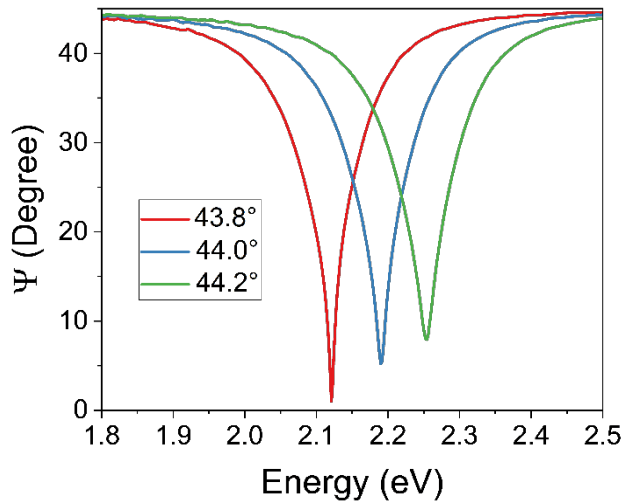

**Figure S2.** Ψ spectra at variable angle of incidence $\theta$ of Ag (51 nm)/$Al_2O_3$ (2 nm).

## S3. Ψ spectra of hybrid stack B at selected angles of incidence

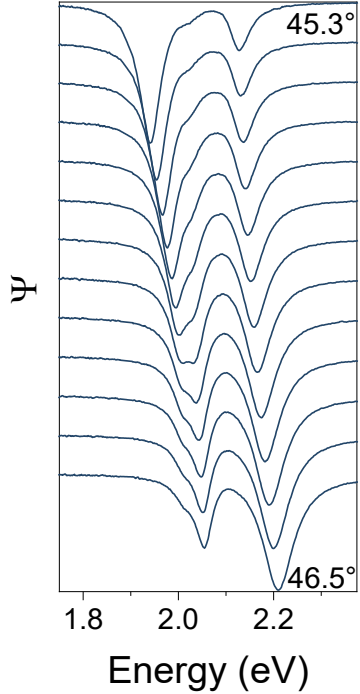

**Figure S3.** Ψ spectra at variable angle of incidence $\theta$ of hybrid stack B.

## S4. Validity of the approximation of negligible WMX-FX coupling

When fitting the experimental dispersion curves with the coupled oscillator model in the main manuscript we neglected direct WMX-FX coupling arguing that the molecules in the TDBC:PVA layers are in average quite far ($\geq 6$ nm) away from the 1L-WS$_2$ surface considering layer thicknesses of 12 nm and 17.5 nm in hybrid stack A and B, respectively. It should be noted that the actual distribution of the molecules in the polymer matrix is not known. Since coupling between WS$_2$ and TDBC excitons may not generally be ruled out, we explore its effect on the polariton dispersion. **Figure S4** shows calculated polariton dispersions with parameters of hybrid stack A including WMX-FX coupling with coupling strengths $g_X = (0-8)$ meV. Apparently, excitonic coupling leads to a repulsion of the upper (UPB) and middle (MPB) polariton branch and an attraction of the MPB and the LPB. . This effect can be counterbalanced in the simulations by increasing the value of $g_{WMX}$ and decreasing of $g_{FX}$ **(see Figure S4b).** Therefore, a fit of the experimental dispersion curves with the coupled oscillator model (eq. 1) cannot provide the information if excitonic coupling is substantial or not since various parameter sets ($g_{WMX}, g_{FX}, g_X$) can reproduce the experimental data. Since the value $g_{WMX} = 25.5$ meV obtained from fits of the experimental dispersion curves with $g_X = 0$ meV is quite similar to a previously reported value[4] we think, however, that it is safe to neglect excitonic coupling in the present sample configuration and note that we might have slightly underestimated $g_{WMX}$ and overestimated $g_{FX}$ in our hybrid stacks by doing so.

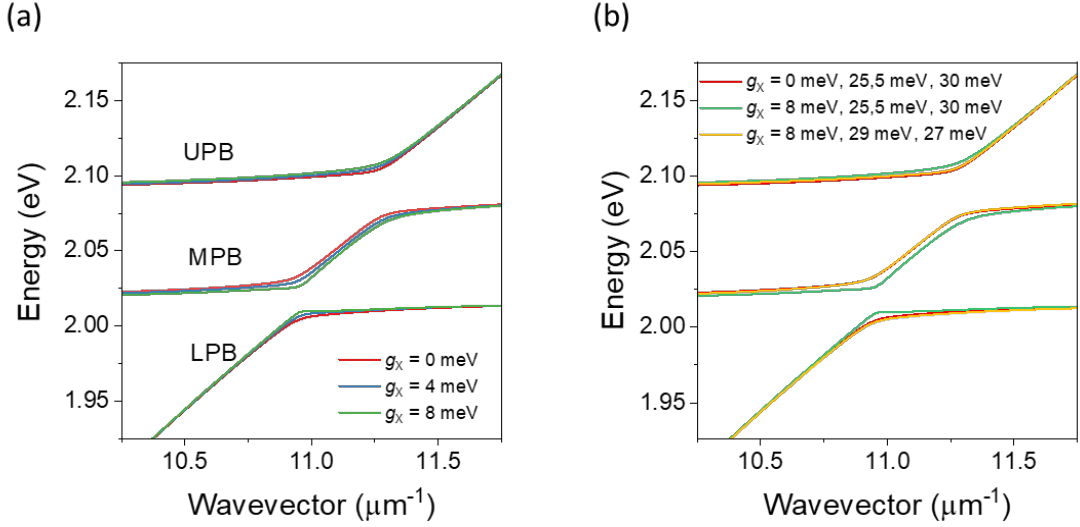

**Figure S4.** Dispersion relation $E$ vs. wavevector $k_\parallel$ of the coupled WMX-FX-SPP resonances calculated with the coupled oscillator model. (a) The parameters are those of hybrid stack A, i.e $g_{WMX} = 25.5$ meV and $g_{FX} = 30$ meV. Included is excitonic coupling between TDBC and 1L-WS$_2$ excitons with a coupling strength $g_X$ given in the legend. (b) Calculation of the dispersion relation modifying all three coupling parameters $g_{WMX}, g_{FX}, g_X$. The values are given in the legend.

**S5. Tuning the Hopfield coefficients of the lower polariton branch**

**Figure S4** compares the plasmonic ($\alpha_P$), WMX ($\alpha_{WMX}$) and FX ($\alpha_{FX}$) contributions (Hopfield coefficients) to the lower polariton branch (LPB) for the dispersions plotted in Figures 3a and b of the main manuscript (i.e. for hybrid stack A and B, respectively) as well as for a hybrid stack with further increased number of TDBC molecules $N_m$ participating in the coupling. By changing $g_{FX} \sim \sqrt{N_m}$, the FX contribution in the LPB can be adjusted. At intermediate wavevector, the excitonic part can be tuned from purely Wannier-Mott exciton-like ($g_{FX} = 0$) to more Frenkel-exciton-like.

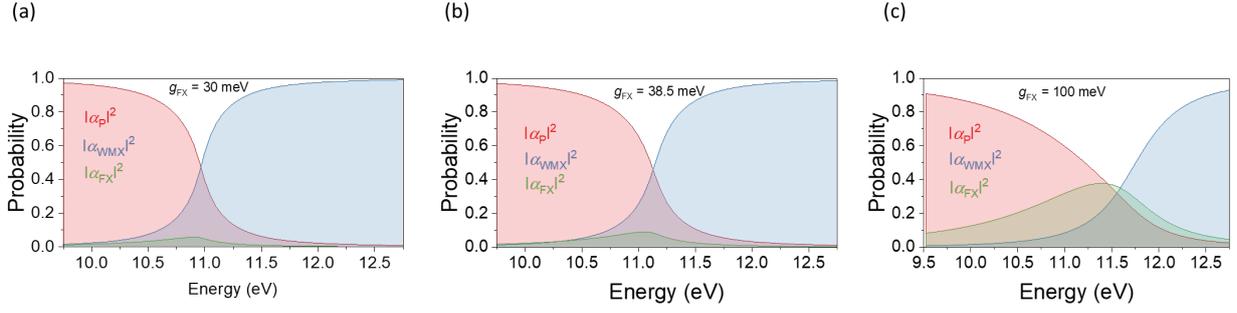

**Figure S5.** Hopfield coefficients of the LPB of hybrid stack A (a) and hybrid stack B (b) extracted from the coupled oscillator model. The parameter $g_{FX}$ is given in the inset, all other parameters are reported in the main manuscript. (c) Hopfield coefficients for the LPB obtained for an increased SPP-Frenkel exciton coupling. The value of $g_{FX}$ corresponds to literature values obtained with TDBC:PVA in the Kretschman-Raether configuration.[4]